\documentclass[sigconf]{acmart}

\AtBeginDocument{%
  \providecommand\BibTeX{{%
    \normalfont B\kern-0.5em{\scshape i\kern-0.25em b}\kern-0.8em\TeX}}}

\setcopyright{acmcopyright}
\copyrightyear{2026}
\acmYear{2026}
\acmDOI{XXXXXXX.XXXXXXX}

\acmConference[CIKM '26]{Proceedings of the 35th ACM International Conference on Information and Knowledge Management}{October 2026}{TBD}
\acmPrice{15.00}
\acmISBN{978-1-4503-XXXX-X/26/10}

\usepackage{booktabs}
\usepackage{graphicx}
\usepackage{amsmath}
\usepackage{algorithm}
\usepackage{algorithmic}
\usepackage{hyperref}

\begin{document}

\title{DSIRM: Learning Query-Bridged Discrete Semantic Identifiers for E-commerce Relevance Modeling}

\author{Bokang Wang}
\affiliation{%
  \institution{Taobao \& Tmall Group of Alibaba}
  \city{Hangzhou}
  \country{China}
}
\email{wangbokang.wbk@alibaba-inc.com}
\email{wangbokang@vip.qq.com}

\author{Xing Fang}
\authornote{Equal contribution as second authors.}
\affiliation{%
  \institution{Taobao \& Tmall Group of Alibaba}
  \city{Hangzhou}
  \country{China}
}
\email{fangxing.fx@taobao.com}

\author{Mingmin Jin}
\authornotemark[1]
\affiliation{%
  \institution{Taobao \& Tmall Group of Alibaba}
  \city{Hangzhou}
  \country{China}
}
\email{jimmy.jmm@taobao.com}

\author{Jing Wang}
\authornote{Corresponding author.}
\affiliation{%
  \institution{Taobao \& Tmall Group of Alibaba}
  \city{Hangzhou}
  \country{China}
}
\email{jing.wangj1@taobao.com}

\author{Zhentao Song}
\affiliation{%
  \institution{Taobao \& Tmall Group of Alibaba}
  \city{Hangzhou}
  \country{China}
}
\email{zhentao.szt@alibaba-inc.com}

\author{Guangxin Song}
\affiliation{%
  \institution{Taobao \& Tmall Group of Alibaba}
  \city{Hangzhou}
  \country{China}
}
\email{guangxin.sgx@taobao.com}

\author{Jianbo Zhu}
\affiliation{%
  \institution{Taobao \& Tmall Group of Alibaba}
  \city{Hangzhou}
  \country{China}
}
\email{zhujianbo.zjb@taobao.com}

\renewcommand{\shortauthors}{Wang, Fang, Jin, et al.}

\begin{abstract}
Despite rapid progress of continuous embeddings for e-commerce search relevance, a long-standing open problem is the difficulty in capturing fine-grained attribute distinctions. While discrete Semantic Identifiers (SIDs) have been widely adopted as a promising alternative, existing SID generation methods rely heavily on unsupervised quantization. In realistic scenarios, the lack of explicit supervision often makes it more difficult to dictate which items should share an SID, resulting in limited capability for query-dependent ranking. To address the issue of unsupervised SIDs, we propose to explicitly model discrete relevance features and develop a Discrete Semantic Identifier Relevance Model (DSIRM). Specifically, we present a query-bridged contrastive quantization approach on the item side, injecting query-item interaction supervision into Residual Quantization to actively learn relevance-aware semantic partitions. On the other hand, we explore generative LLMs on the query side to explicitly predict item SIDs from text, resolving tail queries and intent ambiguity. Hierarchical prefix matching between query and item SIDs yields discriminative features that perfectly complement dense signals. Extensive experimental results on Tmall's production data show that our proposed approach has achieved better results, improving offline AUC by +1.54\%. Deployed via an efficient hybrid architecture, it achieves significant online lifts (+0.13\% UCTR, +0.25\% UCTCVR), proving its massive industrial value.
\end{abstract}



\maketitle

\section{Introduction}
\label{sec:introduction}

As one of the fundamental problems in e-commerce systems, search relevance aims to connect user intent with vast product catalogs. Despite rapid progress in relevance modules, a long-standing open problem is that product titles exhibit extreme homogeneity, which causes substantial lexical overlap across distinct items~\cite{ai2019learning, guo2020deep}. In realistic scenarios, fine-grained attribute distinctions often make it more difficult to discriminate items, resulting in undesired irrelevant results. Furthermore, queries are typically short and ambiguous~\cite{croft2010search, mitra2018introduction}, making relevance strongly query-dependent. Thus, relevance models must simultaneously achieve semantic generalization and fine-grained discrimination~\cite{zhang2020towards, van2016semantic}.

With the prosperity of deep neural networks, dual-encoder architectures~\cite{karpukhin2020dense, xiong2020approximate, humeau2019poly} have been widely adopted as the dominant industrial paradigm, mapping queries and items into continuous embeddings. While these models are capable of strong coarse-grained generalization~\cite{chen2020simple, gao2021simcse}, continuous embeddings face inherent limitations. Items differing only in decisive attributes are densely packed, hindering fine-grained discrimination~\cite{xiong2020approximate, qu2021rocketqa, zhan2021optimizing}. Moreover, a single continuous vector entangles multiple semantic facets, making attribute-level disentanglement difficult for query-dependent judgments~\cite{luan2021sparse, formal2021splade, zamani2018neural}.

Recent work has explored discrete semantic identifiers (SIDs) as an alternative, particularly for generative retrieval~\cite{tay2022transformer, wang2022neural, rajput2023recommender}. SIDs assign hierarchical discrete codes to items, enabling explicit semantic partitioning. However, methods like DSI~\cite{tay2022transformer}, NCI~\cite{wang2022neural}, and TIGER~\cite{rajput2023recommender} primarily target the retrieval stage and rely on unsupervised or self-supervised clustering. Lacking explicit query-item relevance supervision, they struggle to capture the query-dependent distinctions vital for e-commerce ranking.

To address the issue of unsupervised SIDs, we propose to explicitly model the discrete features for query-dependent ranking and develop a novel framework named DSIRM. Specifically, we reposition SIDs from generative retrieval targets to structured relevance features. To relax the constraint of unsupervised clustering, we present an approach to learning hierarchical SIDs through query-bridged contrastive quantization. By leveraging pre-trained dual-tower embeddings and applying InfoNCE loss~\cite{oord2018representation, chen2020simple} directly to query-item pairs within a Residual Quantization (RQ-VAE) process, we guide the quantizer to assign similar SIDs to items co-occurring with similar queries. On the other hand, we fine-tune an autoregressive LLM~\cite{brown2020language, touvron2023llama} to predict item SIDs from text, explicitly resolving intent ambiguity. The resulting hierarchical prefix matching scores complement continuous embeddings in the ranking DNN.

The main contribution of this paper can be summarized as follows:
\begin{itemize}
\item[$\bullet$] Based on the observation that continuous embeddings entangle multiple semantic facets, we propose to reposition SIDs as discrete relevance features specifically designed to augment continuous representations for e-commerce ranking.
\item[$\bullet$] To address the vulnerability of unsupervised clustering in SIDs, we propose a query-bridged contrastive RQ-VAE to actively partition the semantic spaces using query-item interaction signals, leading to a more accurate semantic representation.
\item[$\bullet$] Extensive experimental results on a billion-scale production environment (Tmall) demonstrate that the proposed DSIRM performs better than current state-of-the-art baselines and achieves significant business lifts online.
\end{itemize}

\section{Related Work}
\label{sec:related}

\subsection{Relevance Modeling in E-commerce Search}
In the early stage of relevance modeling, algorithms were mainly based on traditional lexical matching. With the prevailing success of deep learning, dual-encoder architectures~\cite{karpukhin2020dense, xiong2020approximate, humeau2019poly} have been proposed for relevance modeling, where queries and items are independently encoded into a shared continuous embedding space. These learning-based methods achieve strong generalization at the coarse semantic level when trained with contrastive learning~\cite{chen2020simple, gao2021simcse, khattab2020colbert}. However, items differing only in decisive attributes are densely packed in the embedding space, making fine-grained discrimination challenging~\cite{xiong2020approximate, qu2021rocketqa, zhan2021optimizing}. A single continuous vector also entangles multiple semantic facets~\cite{luan2021sparse, formal2021splade}. 

To address these limitations, many works have proposed explicit attribute modeling through multi-field embeddings~\cite{ai2019learning, guo2020deep} or structured knowledge graphs~\cite{sun2019multi, zhang2020semantics}. Other works developed training strategies such as hard negative mining~\cite{xiong2020approximate, zhan2021optimizing, hofstatter2021efficiently} or curriculum learning~\cite{penha2022curriculum, wu2022curriculum} to improve discriminative power. Yet, they remain constrained by the smoothness of the continuous space. Cross-encoder architectures~\cite{nogueira2019passage, khattab2020colbert} achieve stronger matching but incur prohibitive computational costs. These observations motivate exploring complementary discrete representations.

\subsection{Discrete Semantic Identifiers in Retrieval}
Discrete semantic identifiers (SIDs) have emerged as a promising alternative for representing documents and items. DSI~\cite{tay2022transformer} pioneered this paradigm by training transformer models to generate document identifiers conditioned on queries. Following this line of research, NCI~\cite{wang2022neural} proposed hierarchical clustering-based identifiers, and TIGER~\cite{rajput2023recommender} advanced the field by introducing learnable product quantization. More recently, unified semantic identifier systems such as the \textit{One} series~\cite{chen2025onesearch, deng2025onerec, guo2025onesug} have been proposed to generalize across search and recommendation tasks. 

While existing work has demonstrated the effectiveness of SIDs for retrieval, they share a common limitation: the lack of explicit, task-aware supervision during discrete representation learning. Vector quantization approaches rely on unsupervised clustering or self-supervised contrastive learning, failing to address query-dependent relevance. In contrast, our work introduces query-bridged contrastive learning directly into the quantization process, transforming passive discretization into an actively supervised partitioning mechanism.

\section{Preliminaries}
\label{sec:preliminaries}

Relevance modeling in Tmall search is formulated as an ordinal regression task, where a Deep Neural Network (DNN) outputs continuous logits that are subsequently discretized into three relevance grades $y \in \{1, 2, 3\}$. The input features comprise multiple modalities of query-item matching signals: (i) \textbf{dm}: cosine similarity scores computed between query and item continuous text semantic embeddings; (ii) \textbf{mm\_dm}: cosine similarity scores derived from multi-modal embeddings; (iii) \textbf{ct}: discrete category matching scores; and (iv) \textbf{qs}: additional hand-crafted query-side statistical features. The baseline relevance prediction is formulated as:
\begin{equation}
    \text{logit} = \text{DNN}(dm, mm\_dm, ct, qs).
    \label{eq:relevance_compute}
\end{equation}

\begin{figure*}[tbp]
  \centering
  \includegraphics[width=\textwidth]{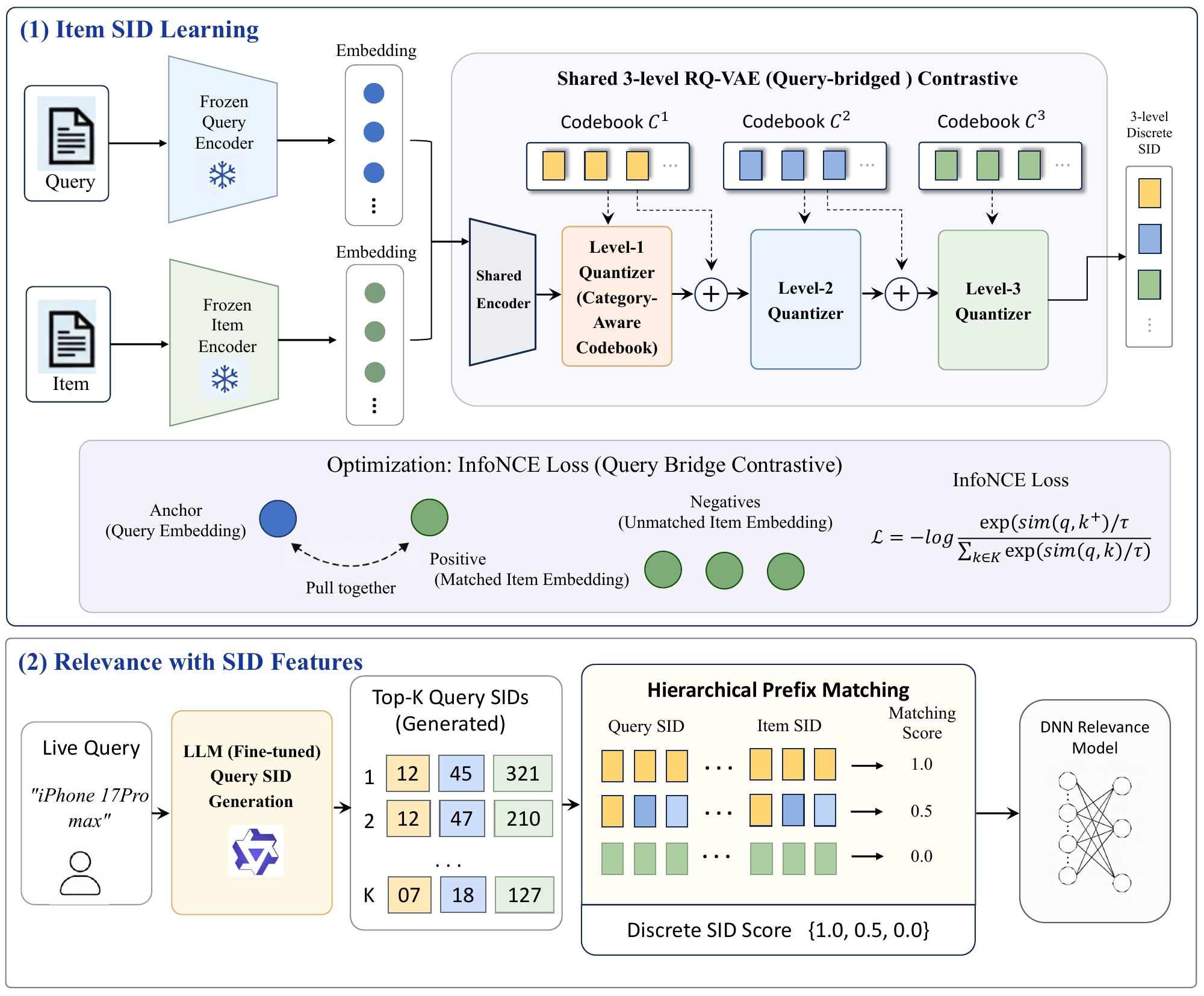}
  \caption{The overall framework of the proposed DSIRM, which models discrete semantic identifiers via query-bridged contrastive quantization. The framework consists of three main components: (1) Item SID learning through contrastive RQ-VAE, (2) Query SID generation via LLM-based prediction, and (3) Hierarchical SID matching for relevance scoring.}
  \label{fig:framework}
\end{figure*}

\section{Methodology} 
\label{sec:methodology}

The general framework of our network, as shown in Figure~\ref{fig:framework}, aims to augment the traditional continuous embedding-based relevance ranking with fine-grained, structured discrete signals. To achieve this, we introduce the SID score (\textbf{ss}), a novel feature computed via hierarchical prefix matching between query and item discrete semantic identifiers. The updated relevance prediction is formulated as:
\begin{equation}
    \text{logit} = \text{DNN}(dm, mm\_dm, ct, qs, ss).
    \label{eq:relevance_compute_new}
\end{equation}

\subsection{Contrastive RQ-VAE for Item SID Learning}
\label{sec:item_sid}

\textbf{Motivation and Limitations of Vanilla RQ-VAE.}
In previous works, the vanilla paradigm proposed by~\cite{lee2022autoregressive} has been widely adopted to train the RQ-VAE. Such a method takes pre-computed embeddings from a fixed model as input and optimizes the quantizer with reconstruction and commitment losses, assuming that semantically similar items will be assigned identical or nearby SIDs. However, such an assumption is too limited. The absence of explicit supervision leaves the clustering behavior entirely dependent on the geometric structure of the input embedding space. Furthermore, the assumption would fail since the notion of item similarity is inherently query-dependent in e-commerce search. Two items may be similar under one query but dissimilar under another, making item-only contrastive learning insufficient.

\textbf{Query-Item Contrastive Learning as Bridge.}
To relax such constraints and provide a more precise calculation of the transmission map (or semantic clustering), we propose to introduce queries as a bridge in the item SID learning process, connecting similar items with the same SID. We construct a dual-tower architecture where both queries and items are first encoded into relevance-aware embeddings through a pre-trained dual-encoder model, then further processed and quantized through the same RQ-VAE. By applying InfoNCE loss on matched query-item pairs, we explicitly guide the quantizer to assign similar SIDs to items that co-occur with the same queries.

\textbf{Dual-Tower Architecture with Shared RQ-VAE.}
Specifically, given a query text $q$ and an item $i$, we first obtain their relevance-aware embeddings from a pre-trained dual-encoder model:
\begin{equation}
\mathbf{e}_q = \text{RelevanceEmb}_q(q), \quad \mathbf{e}_i = \text{RelevanceEmb}_i(i)
\end{equation}
To adapt the pre-trained embeddings to the quantization space, we apply learnable encoders consisting of projection layers:
\begin{equation}
\mathbf{z}_q = \text{Encoder}_q(\mathbf{e}_q), \quad \mathbf{z}_i = \text{Encoder}_i(\mathbf{e}_i),
\end{equation}
where $\text{Encoder}_q$ and $\text{Encoder}_i$ transform the frozen embeddings into a shared latent space $\mathbb{R}^d$. 

We employ a shared hierarchical RQ-VAE with $L$ codebooks $\mathcal{C}_1, \ldots, \mathcal{C}_L$, where $\mathcal{C}_\ell = \{ \mathbf{e}^1_\ell, \ldots, \mathbf{e}^{K_\ell}_\ell \}$ contains $K_\ell$ learnable code vectors. The quantization process iteratively refines the representation:
\begin{equation}
\mathbf{r}_0 = \mathbf{z}, \quad \mathbf{c}_\ell = \arg\min_{\mathbf{e} \in \mathcal{C}_\ell} \|\mathbf{r}_{\ell-1} - \mathbf{e}\|_2^2, \quad \mathbf{r}_\ell = \mathbf{r}_{\ell-1} - \mathbf{c}_\ell.
\end{equation}
The final quantized representation aggregates all levels: $\hat{\mathbf{z}} = \sum_{\ell=1}^{L} \mathbf{c}_\ell$. The discrete SID is the concatenation of codebook indices: $\text{SID} = [k_1, k_2, \ldots, k_L]$, where $k_\ell \in \{1, \ldots, K_\ell\}$. 

Our training objective combines three complementary losses. First, we employ a symmetric InfoNCE loss to align query-item pairs:
\begin{equation}
\mathcal{L}_{\text{InfoNCE}} = \frac{1}{2}\left(\mathcal{L}_{q \to i} + \mathcal{L}_{i \to q}\right),
\end{equation}
\begin{equation}
\mathcal{L}_{q \to i} = -\frac{1}{B}\sum_{m=1}^{B} \log \frac{\exp(\text{sim}(\hat{\mathbf{z}}_q^m, \hat{\mathbf{z}}_i^m) / \tau)}{\sum_{n=1}^{B} \exp(\text{sim}(\hat{\mathbf{z}}_q^n, \hat{\mathbf{z}}_i^n) / \tau)},
\end{equation}
\begin{equation}
\mathcal{L}_{i \to q} = -\frac{1}{B}\sum_{m=1}^{B} \log \frac{\exp(\text{sim}(\hat{\mathbf{z}}_i^m, \hat{\mathbf{z}}_q^m) / \tau)}{\sum_{n=1}^{B} \exp(\text{sim}(\hat{\mathbf{z}}_i^n, \hat{\mathbf{z}}_q^n) / \tau)}.
\end{equation}
Second, to stabilize the hierarchical codebook learning, we apply a commitment loss across all $L$ levels:
\begin{equation}
\mathcal{L}_{\text{commit}} = \frac{1}{L}\sum_{\ell=1}^{L} \sum_{x \in \{q, i\}} \|\mathbf{z}_x - \text{sg}[\hat{\mathbf{z}}_x^{(\ell)}]\|_2^2,
\end{equation}
where $\hat{\mathbf{z}}_x^{(\ell)}$ denotes the partial quantized sum up to level $\ell$. Third, we reconstruct the continuous embeddings using lightweight shared decoders $\mathcal{D}$:
\begin{equation}
\mathcal{L}_{\text{recon}} = \sum_{x \in \{q, i\}} \|\mathcal{D}(\hat{\mathbf{z}}_x) - \mathbf{z}_x\|_2^2.
\end{equation}
The final joint objective is defined as:
\begin{equation}
\mathcal{L} = \lambda_{\text{InfoNCE}}\mathcal{L}_{\text{InfoNCE}} + \lambda_{\text{commit}}\mathcal{L}_{\text{commit}} + \lambda_{\text{recon}} \mathcal{L}_{\text{recon}}.
\label{eq:weight}
\end{equation}

\textbf{Category-Aware Codebook Allocation.}
E-commerce catalogs exhibit extreme category imbalance. Without explicit category separation at the first quantization level, the RQ-VAE learning would be dominated by high-frequency categories, resulting in weak representations for tail categories. To address this issue, we propose to explicitly encode category structure into the first-level codebook. Specifically, we maintain a category mapping $\phi: \mathcal{C}_{\text{cat}} \to \{1, \ldots, K_1\}$ that assigns each category to a pre-defined code in the first codebook $\mathcal{C}_1$. During quantization, items from category $c$ are constrained to use code $\phi(c)$:
\begin{equation}
\mathbf{c}_1 = \mathcal{C}_1[\phi(c)].
\end{equation}
The category-aware first-level codebook is updated via exponential moving average (EMA):
\begin{equation}
\mathcal{C}_1[\phi(c)] \leftarrow \alpha \cdot \mathcal{C}_1[\phi(c)] + (1-\alpha) \cdot \mathbb{E}_{i \in c}[\mathbf{z}_i],
\end{equation}
where $\alpha$ is the decay rate and $\mathbb{E}_{i \in c}[\mathbf{z}_i]$ is the mean representation of items in category $c$.

\subsection{Generative LLM for Query SID Learning}
\label{sec:query_sid}

\textbf{Motivation: Generative Query Understanding.}
Our generative design is fundamentally motivated by two inherent challenges in e-commerce query understanding. On the one hand, e-commerce search frequently encounters exploratory queries that necessitate external commonsense reasoning beyond lexical matching. Pre-trained LLMs inherently encode extensive world knowledge, enhancing generalization capabilities. On the other hand, queries are typically short and highly ambiguous. An auto-regressive LLM can generate a diverse distribution of potential semantic identifiers, explicitly modeling the multifaceted nature of user intent.

\textbf{Supervised Fine-Tuning for SID Prediction.}
We adopt a standard supervised fine-tuning (SFT) paradigm to train an auto-regressive LLM to predict item SIDs conditioned on query text. The training data consists of query-target pairs extracted from historical user interaction logs. For each interaction between a query $q$ and an item $i$, we retrieve the item's hierarchical SID $\text{SID}_i = [k_1, k_2, \ldots, k_L]$ previously assigned by the RQ-VAE model. The LLM is trained to minimize the standard auto-regressive cross-entropy loss:
\begin{equation}
\mathcal{L}_{\text{SFT}} = -\sum_{t=1}^{L} \log P(k_t | q, k_{<t}; \theta).
\end{equation}

\textbf{Inference via Multi-Cluster Decoding.}
At inference time, to explicitly materialize the modeled intent ambiguity, we approximate the conditional distribution of relevant item clusters by employing beam search decoding. The model generates the top-$K$ most probable SID sequences:
\begin{equation}
\{\text{SID}_1, \text{SID}_2, \ldots, \text{SID}_K\} = \arg\max_{\text{SID}} P(\text{SID} | q; \theta).
\end{equation}
By generating multiple SIDs, the model can simultaneously capture different facets of query intent.

\subsection{SID-based Relevance Scoring and Ranking}
\label{sec:dnn}

\textbf{Hierarchical SID Matching.}
Given a query $q$ with generated SIDs $\{\text{SID}_q^1, \ldots, \text{SID}_q^K\}$ and an item $i$ with its learned SID $\text{SID}_i = [k_1^i, k_2^i, \ldots, k_L^i]$, we compute the matching score based on hierarchical prefix matching. Matching at deeper levels indicates stronger semantic alignment. The SID matching algorithm proceeds in a cascading manner. Taking $L = 3$ as an example, Algorithm~\ref{alg:sid_matching} presents the detailed procedure.

\begin{algorithm}[h]
\caption{Hierarchical SID Matching}
\label{alg:sid_matching}
\begin{algorithmic}[1]
\REQUIRE Query SID list = $\{\text{SID}_q^1, \ldots, \text{SID}_q^K\}$, Item SID $\text{SID}_i = [k_1^i, k_2^i, k_3^i]$
\ENSURE SID matching score $\text{Score}_{\text{SID}}(q, i)$
\STATE $\text{level} \gets 0$ \COMMENT{Maximum matching level}
\FOR{$j = 1$ to $K$}
  \STATE $\ell \gets 0$ \COMMENT{Current matching depth for $\text{SID}_q^j$}
  \IF{$k_1^{q,j} = k_1^i$}
    \STATE $\ell \gets 1$
    \IF{$k_2^{q,j} = k_2^i$}
      \STATE $\ell \gets 2$
      \IF{$k_3^{q,j} = k_3^i$}
        \STATE $\ell \gets 3$
      \ENDIF
    \ENDIF
  \ENDIF
  \STATE $\text{level} \gets \max(\text{level}, \ell)$
  \IF{$\text{level} = 3$}
    \STATE \textbf{break} \COMMENT{Early termination at full match}
  \ENDIF
\ENDFOR
\RETURN $\text{level}$
\end{algorithmic}
\end{algorithm}

The SID score (ss) is assigned based on the deepest matching level:
\begin{equation}
ss(q, i) = 
\begin{cases}
0.0   & \text{if level} = 0, \\
0.25  & \text{if level} = 1, \\
0.5   & \text{if level} = 2, \\
1.0   & \text{if level} = 3.
\end{cases}
\label{eq:sid_score}
\end{equation}

\textbf{Integration into Relevance Ranking DNN.}
The SID score is incorporated as an additional feature into a deep neural network (DNN) model. The model processes these features through an embedding layer for discrete features and concatenates them with continuous features:
\begin{equation}
\mathbf{x}(q, i) = [\text{dm}, \text{mm\_dm}, \text{ct}, \text{ss}, \text{qs}].
\end{equation}
The relevance logit is computed through a multi-layer perceptron: $\text{logit}(q, i) = \text{MLP}(\mathbf{x}(q, i))$. 

\section{Experimental Results}
\label{sec:experiments}

\subsection{Experimental Setups}

\textbf{Datasets and Metrics.}
Both SID learning datasets and relevance scoring datasets are used to evaluate our proposed method. For SID learning, we use large-scale real-world data from Tmall, containing approximately 80 million query-item pairs $(q, i)$. For relevance scoring, we construct a labeled dataset generated entirely by Qwen3-30B using carefully designed prompts. The validation subset achieved an exceptionally high consistency rate of approximately 94\% against human judgments, demonstrating that the LLM-derived labels are highly credible. The final training set comprises approximately 1.6 million labeled pairs, while the test set contains 100,000 pairs. To compare our method with state-of-the-art baselines, the widely used metrics Precision, Recall, and Area Under the ROC Curve (AUC) are adopted.

\textbf{Implementation Details.}
For Item SID Learning, we use pre-trained multi-modal relevance embeddings as input, and a hierarchical RQ-VAE with $K_1 = 216$, $K_2 = 512$, and $K_3 = 512$. Notably, $K_1 = 216$ corresponds exactly to the number of primary (1st-level) categories in the Tmall e-commerce product catalog, which strictly aligns with our category-aware codebook allocation strategy. The network is optimized using batch size 256, temperature $\tau = 0.07$, $\lambda_{\text{InfoNCE}} = 1.0$, $\lambda_{\text{recon}} = 0.1$, and $\lambda_{\text{commit}} = 1.0$. For Query SID Learning, we fine-tune Qwen3-0.6B~\cite{yang2025qwen3} with batch size 128. Inference uses beam search with $K = 5$. The DNN relevance model is implemented as a 3-layer MLP. 

To ensure a fair comparison with the standard retrieval-oriented SID baselines (DSI~\cite{tay2022transformer} and TIGER~\cite{rajput2023recommender}), we solely employ their proposed methodologies for generating the item SIDs. Subsequently, we apply the exact same query-side LLM generator and the identical downstream DNN relevance scoring architecture (as formulated in Eq.~\ref{eq:relevance_compute_new}) to evaluate them. This experimental setup isolates the variables, ensuring that performance differences strictly reflect the quality of the generated item SIDs rather than variations in query encoding or scoring architectures.

\subsection{Overall Performance}

In Table~\ref{tab:main_results}, we compare our performance with some state-of-the-art methods on the offline dataset. We investigate the fundamental design choice of input representation for RQ-VAE: static pre-trained embeddings (DSIRM) versus dynamic encoding (exp0, exp1), as well as comparison against standard retrieval-oriented SID baselines (DSI~\cite{tay2022transformer} and TIGER~\cite{rajput2023recommender}). 

\begin{table}[h]
\centering
\caption{Comparison against various SID generation methods and input representation strategies. Precision and Recall are computed at threshold 0.5.}
\label{tab:main_results}
\small
\begin{tabular}{lcccc}
\toprule
\textbf{Method} & \textbf{Input} & \textbf{Prec. / Rec.} & \textbf{Prec. / Rec.} & \textbf{AUC} \\
 & \textbf{Type} & \textbf{(Positive)} & \textbf{(Negative)} & \\
\midrule
Baseline (w/o SID) & - & 87.00 / 90.00 & 81.40 / 76.50 & 0.9202 \\
\midrule
DSI~\cite{tay2022transformer} & Dynamic & 85.73 / 92.30 & 84.58 / 73.33 & 0.9312 \\
TIGER~\cite{rajput2023recommender} & Static & 85.37 / 93.54 & 85.42 / 70.24 & 0.9323 \\
\midrule
exp0 (mT5) & Dynamic & 85.85 / 92.36 & 84.47 / 73.17 & 0.9308 \\
exp1 (BERT) & Dynamic & 85.90 / 92.29 & 84.37 / 73.32 & 0.9309 \\
\textbf{exp2 (DSIRM)} & Static & 85.57 / \textbf{93.62} & \textbf{86.39} / 71.97 & \textbf{0.9356} \\
\bottomrule
\end{tabular}
\end{table}

Our method achieves the best performance with 0.9356 AUC, outperforming the baseline without SIDs and all competing SID approaches. While DSI and TIGER rely on generative and clustering-based quantization paradigms respectively, their lack of explicit query-item interaction supervision restricts their clustering behavior. In contrast, our proposed approach actively partitions the semantic space based on search relevance. Notably, static embeddings substantially outperform dynamic encoding. Decoupling representation learning from quantization allows the RQ-VAE to focus exclusively on discovering hierarchical discrete structures.

\subsection{Ablation Study}

\textbf{Ablation study of different components.}
To explore the effectiveness of the different elements in our method, we performed an ablation study to analyze the contribution of each. Table~\ref{tab:ablation_components} presents the results. Removing contrastive learning (exp6) results in -0.09\% AUC degradation on static embeddings, and removing category constraints (exp5) leads to a -0.07\% AUC drop. The results prove that query-bridged contrastive learning successfully transforms passive clustering into active relevance-aware partitioning, and category constraints are crucial for maintaining tail category representations.

\begin{table}[h]
\centering
\caption{Ablation study on proposed components.}
\label{tab:ablation_components}
\small
\begin{tabular}{lcccc}
\toprule
\textbf{Method} & \textbf{Contr.} & \textbf{Category} & \textbf{Prec. / Rec.} & \textbf{AUC} \\
 & \textbf{Learning} & \textbf{Constraint} & \textbf{(Positive)} & \\
\midrule
\multicolumn{5}{l}{\textit{Dynamic Encoding (BERT)}} \\
\midrule
exp1 (Full) & \checkmark & \checkmark & 85.90 / 92.29 & 0.9309 \\
exp3 (w/o Cat.) & \checkmark & $\times$ & 85.96 / 92.21 & 0.9300 \\
exp4 (w/o Contr.) & $\times$ & \checkmark & 84.28 / 93.80 & 0.9288 \\
\midrule
\multicolumn{5}{l}{\textit{Static Pre-trained Embeddings}} \\
\midrule
\textbf{exp2 (DSIRM)} & \checkmark & \checkmark & 85.57 / \textbf{93.62} & \textbf{0.9356} \\
exp5 (w/o Cat.) & \checkmark & $\times$ & 85.64 / 93.42 & 0.9349 \\
exp6 (w/o Contr.) & $\times$ & \checkmark & 85.57 / 93.51 & 0.9347 \\
\bottomrule
\end{tabular}
\end{table}

\textbf{Ablation study of different weights in the loss function.}
We analyze sensitivity to loss weight hyperparameters $\lambda_{\text{recon}}$ and $\lambda_{\text{commit}}$ on the static embedding configuration. As shown in Table~\ref{tab:ablation_loss_weights}, lower reconstruction weight ($\lambda_{\text{recon}} = 0.1$) outperforms higher values. When leveraging high-quality pre-trained embeddings, exact reconstruction becomes less critical than learning discriminative discrete codes. Higher commitment weight ($\lambda_{\text{commit}} = 1.0$) ensures stable codebook learning.

\begin{table}[h]
\centering
\caption{Hyperparameter sensitivity analysis on static embeddings.}
\label{tab:ablation_loss_weights}
\small
\begin{tabular}{ccccc}
\toprule
$\lambda_{\text{InfoNCE}}$ & $\lambda_{\text{recon}}$ & $\lambda_{\text{commit}}$ & \textbf{Prec. / Rec.} & \textbf{AUC} \\
 & & & \textbf{(Positive)} & \\
\midrule
1.0 & 0.1 & 1.0 & 85.57 / \textbf{93.62} & \textbf{0.9356} \\
1.0 & 1.0 & 0.25 & 85.65 / 93.50 & 0.9354 \\
1.0 & 0.1 & 0.25 & 86.04 / 93.26 & 0.9348 \\
1.0 & 1.0 & 1.0 & 85.85 / 93.26 & 0.9350 \\
\bottomrule
\end{tabular}
\end{table}

\subsection{Online Deployment and Performance}
\label{sec:online_deployment}

\textbf{Offline-Online Hybrid Serving Architecture.} 
Deploying Generative LLMs and complex discrete matching logic into a billion-scale search engine poses severe latency constraints. Therefore, we implement an efficient offline-online hybrid deployment architecture. Item SIDs are pre-computed offline utilizing the trained RQ-VAE model and stored in a distributed in-memory KV database. Fetching an item's SID requires only an $O(1)$ memory lookup. Crucially, for newly added items, we apply a fallback mechanism by assigning a default feature value of $ss=0.0$, ensuring they can still be robustly retrieved based on traditional dense matching features. Furthermore, to balance computational cost and latency on the query side, we use a tiered routing mechanism. Approximately $88\%$ of the total search traffic hits an offline cache that pre-generates SIDs for historical queries. For the remaining $12\%$ of real-time online traffic, we deploy an online inference service. The service achieves an average response time (RT) of 17.9ms and a P99 RT of 24.3ms, introducing almost zero latency overhead to the primary ranking pathway.

\textbf{Online A/B Testing Results.} 
Over the continuous online evaluation period on Tmall, DSIRM yielded highly significant business improvements over the continuous matching baseline:
\begin{itemize}
    \item \textbf{+0.13\% User Click-Through Rate (UCTR)}
    \item \textbf{+0.25\% User Click-To-Conversion Rate (UCTCVR)}
\end{itemize}
These results compellingly validate that repositioning discrete semantic identifiers from generative retrieval to structured relevance features is a highly effective, cost-efficient, and scalable paradigm.

\section{Conclusion}
\label{sec:conclusion}

In this paper, we analyze the problem of continuous embeddings in e-commerce search and propose to reposition semantic identifiers from generative retrieval targets to structured relevance features. To address the challenge with unsupervised quantization in SID generation, we have developed a query-bridged contrastive quantization method that actively partitions the semantic space based on search relevance. Moreover, we introduce an auto-regressive LLM to handle complex intent ambiguity. We have comprehensively evaluated our proposed method on both offline industrial datasets and online production environments. The result proves the effectiveness and massive industrial value of our method.

\section*{Generative AI Usage Disclosure}
\label{sec:genai_disclosure}
In accordance with ACM's policy, the authors disclose that large language models (LLMs) were used during the preparation of this manuscript for proofreading and grammatical correction. The generation of core technical ideas, algorithm designs, experimental setups, and data analyses were entirely conducted by the human authors.

\section*{Acknowledgments}
This work was supported by the Taobao \& Tmall Group of Alibaba. We thank the search relevance team for their valuable feedback and infrastructure support.

\bibliographystyle{ACM-Reference-Format}
\bibliography{references}

\end{document}